\documentclass[conference]{IEEEtran}
\IEEEoverridecommandlockouts
\usepackage{cite}
\usepackage{amsmath,amssymb,amsfonts}
\usepackage{algorithmic}
\usepackage{algorithm}
\usepackage{graphicx}
\usepackage{textcomp}
\usepackage{xcolor}
\def\BibTeX{{\rm B\kern-.05em{\sc i\kern-.025em b}\kern-.08em
    T\kern-.1667em\lower.7ex\hbox{E}\kern-.125emX}}

\renewcommand{\baselinestretch}{0.98}

\begin{document}

\title{\huge Movable Antenna Enhanced AF Relaying: Two-Stage\\ Antenna Position Optimization}

\author{
	\IEEEauthorblockN{Nianzu Li\IEEEauthorrefmark{2},~Weidong Mei\IEEEauthorrefmark{4},~Boyu Ning\IEEEauthorrefmark{4}, and Peiran Wu\IEEEauthorrefmark{2}}
	\IEEEauthorblockA{\IEEEauthorrefmark{2}School of Electronics and Information Technology, Sun Yat-sen University, Guangzhou, China 510006.}
	\IEEEauthorblockA{\IEEEauthorrefmark{4}National Key Laboratory of Wireless Communications, \\University of Electronic Science and Technology of China, Chengdu, China 611731.}
	\IEEEauthorblockA{E-mail: linz5@mail2.sysu.edu.cn, wmei@uestc.edu.cn, boydning@outlook.com, wupr3@mail.sysu.edu.cn}
}

\maketitle

\begin{abstract}
The movable antenna (MA) technology has attracted increasing attention in wireless communications due to its capability for flexibly adjusting the positions of multiple antennas in a local region to reconfigure channel conditions. In this paper, we investigate its application in an amplify-and-forward (AF) relay system, where a multi-MA AF relay is deployed to assist in the wireless communications from a source to a destination. In particular, we aim to maximize the achievable rate at the destination, by jointly optimizing the AF weight matrix at the relay and its MAs' positions in two stages for receiving the signal from the source and transmitting its amplified version to the destination, respectively. However, compared to the existing one-stage antenna position optimization, the two-stage position optimization is more challenging due to its intricate coupling in the achievable rate at the destination. To tackle this challenge, we decompose the considered problem into several subproblems by invoking the alternating optimization (AO) and solve them by using the semidefinite programming and the gradient ascent. Numerical results demonstrate the superiority of our proposed system over the conventional relaying system with fixed-position antennas (FPAs) and also drive essential insights.
\end{abstract}


\section{Introduction}
With the evolution of wireless communication systems, multiple-input multiple-output (MIMO) and massive MIMO technologies have been widely promoted and investigated in both academia and industry to pursue higher data rate and larger capacity. However, by only relying on coventional fixed-position antennas (FPAs), the spatial variation of wireless channels cannot be fully exploited, which may result in suboptimal communication performance\cite{ref1}. Furthermore, high energy consumption and hardware cost remain critical issues due to the increasing number of antennas and radio frequency (RF) chains, especially in a high frequency band. To overcome these limitations, the movable antenna (MA) technology was recently proposed as a promising
solution, which allows for flexible antenna movement within a local region and thereby provides additional degrees of freedom (DoFs) to improve the communication performance without the need for increasing the number of antennas\cite{ref1},\cite{ref3},\cite{ref35}. 

Motivated by the promising benefits of MAs, some recent works have investigated their position optimization problems under different scenarios. For example, the authors in \cite{ref4} aimed to maximize the channel capacity of an MA-enhanced MIMO communication system by jointly optimizing the transmit covariance matrix and the positions of MAs in both transmit and receive regions. In \cite{ref9}, the authors investigated the uplink of an MA-aided multi-user communication system, aiming to minimize the total transmit power of multiple users by jointly optimizing the positions of their equipped MAs. The MA-enhanced non-orthogonal multiple access (NOMA) was studied in \cite{ref2}, where a low-complexity algorithm was proposed to maximize the system's sum rate. Unlike the above works optimizing the positions of MAs in a continuous space, the authors in \cite{ref27} discretized this space into a multitude of sampling points and proposed a novel graph-based algorithm to select an optimal subset of the sampling points for maximizing the achievable rate of a multiple-input single-output (MISO) system. In addition, MAs have also been applied in other system setups, such as physical-layer security\cite{ref5}, cognitive radio\cite{ref29}, and intelligent reflecting surface (IRS)-aided communication systems\cite{ref34}.

\begin{figure}[t]
	\centering
	\includegraphics[width=0.45\textwidth]{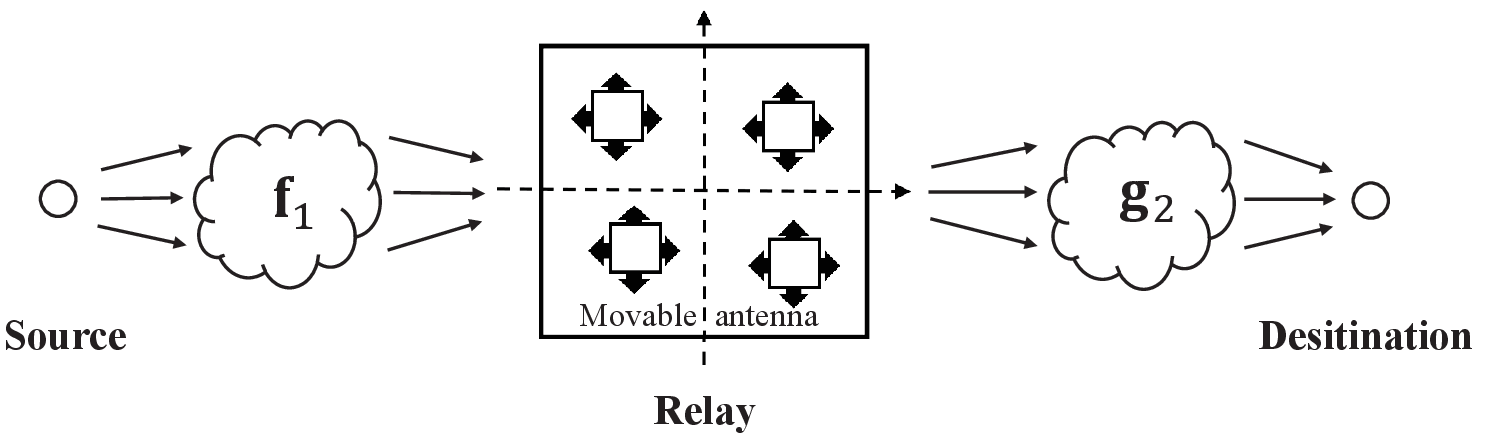}
	\caption{MA-enhanced AF relay system.}
	\label{system_model}
	\vspace{-0.25cm}
\end{figure}
However, MAs have not been applied to a relay system so far in the literature, while they can be exploited to assist in both the information reception and transmission of a relay. To fill in this gap, we investigate in this paper an MA-enhanced amplify-and-forward (AF) relaying system, where a single-antenna source aims to communicate with a single-antenna destination with the aid of a multi-MA relay, as shown in Fig \ref{system_model}. In particular, we aim to maximize the achievable rate at the destination by jointly optimizing the AF weight matrix at the relay and its MAs' positions in two stages, for receiving the signal from the source and transmitting its amplified version to the destination, respectively. However, compared to the one-stage antenna position optimization as studied in the existing works, the two-stage antenna position optimization appears to be more challenging to tackle due to its intricate coupling in the achievable rate at the destination. To resolve this issue, we decouple the joint optimization problem into several subproblems by invoking an alternating optimization (AO) framework and solve them separately by combining the semidefinite programming and the gradient ascent (GA). Simulation results show that our proposed scheme can significantly outperform the conventional FPA-based relaying scheme and drive essential insights into the MA position optimization.

\textit{Notations}: Boldface lowercase and uppercase letters denote vectors and matrices, respectively. $(\cdot)^T$, $(\cdot)^{*}$, and $(\cdot)^H$ denote transpose, conjugate, and conjugate transpose, respectively. $\mathcal{CN}(0,\mathbf{\Sigma})$ denotes the circularly symmetric complex Gaussian distribution with zero mean and covariance matrix $\mathbf{\Sigma}$. $\|\mathbf{A}\|$, $\mathrm{rank}(\mathbf{A})$, and $\mathrm{Tr}(\mathbf{A})$ denote the Frobenius norm, rank, and trace of the matrix $\mathbf{A}$, respectively. $\mathrm{vec}(\mathbf{A})$ denotes the vectorization of the matrix $\mathbf{A}$. $\otimes$ denotes the Kronecker product.

\section{System Model}
As shown in Fig. \ref{system_model}, we consider an AF relay system, which consists of a source node, a multi-antenna relay, and a destination node. The relay is equipped with $N$ MAs, while each of the other nodes is equipped with a single FPA\footnote{This can practically occur in device-to device communications, where remote sensors aim to communicate with each other via multi-hop transmission.}. We assume that the positions of the $N$ MAs can be flexibly adjusted within a two-dimensional (2D) region ${\cal C}_r$ of size $A \times A$, where $A$ denotes the length of the region per dimension. We consider a challenging case where the direct link between the source and the destination is severely blocked due to the dense obstacles in the environment. Thus, an AF relay with $N$ MAs is deployed to assist in their direct communications. We assume that the relay operates in the half-duplex mode, which divides the source-destination communication into two stages. In the first stage, the source transmits its signal to the relay, and its received signal is given by
\begin{equation}
	\mathbf{y}_r=\sqrt{P_s}\mathbf{h}_1 s+\mathbf{n}_r,
\end{equation}
where $\mathbf{h}_1\in\mathbb{C}^{N\times 1}$ is the channel vector from the source to the relay, $s$ is the transmitted signal from the source with $\mathbb{E}[|s|^2]=1$, $P_s$ is the transmit power, and $\mathbf{n}_r\sim\mathcal{CN}(\mathbf{0},\sigma_r^2\mathbf{I}_N)$ is the additive white Gaussian noise (AWGN) at the relay. In the second stage, the relay processes the received signal $\mathbf{y}_r$ by an AF weight matrix $\mathbf{W}\in\mathbb{C}^{N\times N}$, and then forwards it to the destination. Therefore, the received signal at the destination is given by
\begin{equation}
	y_d=\sqrt{P_s}\mathbf{h}_2^H\mathbf{W}\mathbf{h}_1s+\mathbf{h}_2^H\mathbf{W}\mathbf{n}_r+n_d,
\end{equation}
where $\mathbf{h}_2\in\mathbb{C}^{N\times 1}$ is the channel vector from the relay to the destination, and $n_d\sim\mathcal{CN}(0,\sigma_d^2)$ is the AWGN at the destination.


To facilitate the information reception/transmission from/to the source/destination, the positions of the $N$ MAs at the relay need to be adjusted twice at the beginning of the two stages. Accordingly, let $\tilde{\mathbf{r}}=[\mathbf{r}_1,\cdots,\mathbf{r}_N]\in\mathbb{C}^{2\times N}$ and $\tilde{\mathbf{t}}=[\mathbf{t}_1,\cdots,\mathbf{t}_N]\in\mathbb{C}^{2\times N}$ denote the collections of the $N$ MAs' positions in the first and second stages, respectively, where $\mathbf{r}_n=[x_{r,n},y_{r,n}]^T\in\mathcal{C}_r$ and $\mathbf{t}_n=[x_{t,n},y_{t,n}]^T\in\mathcal{C}_r,\forall n$. By applying a similar field-response channel model as in\cite{ref3}, the channel from the source to the relay in the first stage can be represented as
\begin{equation}
	\mathbf{h}_1(\tilde{\mathbf{r}})=\mathbf{F}_1^H(\tilde{\mathbf{r}})\mathbf{g}_1=\left[\mathbf{f}_1(\mathbf{r}_1),\cdots,\mathbf{f}_1(\mathbf{r}_N)\right]^H\mathbf{g}_1,
\end{equation}
where 
\begin{equation}
	\mathbf{f}_1(\mathbf{r}_n)=\left[e^{\mathrm{j}\frac{2\pi}{\lambda}\rho_{1,1}(\mathbf{r}_n)},\cdots,e^{\mathrm{j}\frac{2\pi}{\lambda}\rho_{1,L_r}(\mathbf{r}_n)}\right]^T,\forall n,
\end{equation}
is the receive field-response vector (FRV) at the relay with $\rho_{1,i}(\mathbf{r}_n)=x_{r,n}\sin\theta_{1,i}\cos\phi_{1,i}+y_{r,n}\cos\theta_{1,i},1\leq i \leq L_r$, and $\mathbf{g}_1$ is the path-response vector (PRV) from the source to the relay. Here, $\lambda$ is the  carrier wavelength, $\theta_{1,i}$ and $\phi_{1,i}$ are the elevation and azimuth angles of arrival (AoAs) of the $i$-th receive path from the source to the relay, respectively. Similarly, in the second stage, the channel from the relay to the destination is given by
\begin{equation}
	\mathbf{h}_2(\tilde{\mathbf{t}})=\mathbf{G}_2^H(\tilde{\mathbf{t}})\mathbf{f}_2=\left[\mathbf{g}_2(\mathbf{t}_1),\cdots,\mathbf{g}_2(\mathbf{t}_N)\right]^H\mathbf{f}_2,
\end{equation}
where 
\begin{equation}
	\mathbf{g}_2(\mathbf{t}_n)=\left[e^{\mathrm{j}\frac{2\pi}{\lambda}\rho_{2,1}(\mathbf{t}_n)},\cdots,e^{\mathrm{j}\frac{2\pi}{\lambda}\rho_{2,L_t}(\mathbf{t}_n)}\right]^T,\forall n,
\end{equation}
denotes the transmit FRV at the relay with $\rho_{2,i}(\mathbf{t}_n)=x_{t,n}\sin\theta_{2,i}\cos\phi_{2,i}+y_{t,n}\cos\theta_{2,i},1\leq i \leq L_t$. Here, $\mathbf{f}_2$, $\theta_{2,i}$ and $\phi_{2,i}$ denote the PRV, elevation and azimuth angles of departure (AoDs) of the $i$-th transmit path from the relay to the destination, respectively.

Based on the above, the end-to-end signal-to-noise ratio (SNR) at the destination can be expressed as 
\begin{equation}
	\label{eq7}
	\gamma=\frac{P_s\left|\mathbf{h}_2^H(\tilde{\mathbf{t}})\mathbf{W}\mathbf{h}_1(\tilde{\mathbf{r}})\right|^2}{\sigma_r^2\left\|\mathbf{h}_2^H(\tilde{\mathbf{t}})\mathbf{W}\right\|^2+\sigma_d^2},
\end{equation}
which is a function of $\tilde{\mathbf{r}},\tilde{\mathbf{t}}$ and $\mathbf{W}$. Our goal is to maximize \eqref{eq7} by jontly optimizing $\tilde{\mathbf{r}},\tilde{\mathbf{t}}$ and $\mathbf{W}$. Accordingly, the optimization problem is formulated as 
\begin{subequations}
	\label{eq8}
	\begin{align}
		\text{(P1)} \quad \max_{\mathbf{W},\tilde{\mathbf{r}},\tilde{\mathbf{t}}}& \quad \gamma\\
		\mathrm{s.t.}
		& \quad \tilde{\mathbf{r}}\in\mathcal{C}_r,\tilde{\mathbf{t}}\in\mathcal{C}_r,\label{eq8b}\\
		& \quad \|\mathbf{r}_m-\mathbf{r}_n\|_2\geq D,~1\leq m\neq n\leq N,\label{eq8c}\\
		& \quad \|\mathbf{t}_m-\mathbf{t}_n\|_2\geq D,~1\leq m\neq n\leq N,\label{eq8d}\\
		& \quad 
		P_s\|\mathbf{W}\mathbf{h}_1(\tilde{\mathbf{r}})\|^2+\sigma_r^2\|\mathbf{W}\|^2\leq P_{\mathrm{tot}},
	\end{align}
\end{subequations}
where $D$ is the minimum inter-MA distance at the relay to avoid mutual coupling, and $P_{\mathrm{tot}}$ is the total power budget of the relay. Note that to investigate the fundamental limit of an MA-assisted relay system, we assume that all involved channel state information (CSI), i.e., $\mathbf{h}_1(\tilde{\mathbf{r}})$'s and $\mathbf{h}_2(\tilde{\mathbf{t}})$'s, is available. In practice, this can be achieved by applying some existing channel estimation techniques designed for MAs \cite{ref6}. However, (P1) is non-convex due to its non-concave objective function and the intricate coupling of $\tilde{\mathbf{r}}$ and $\tilde{\mathbf{t}}$ therein. In the next section, we will apply an AO algorithm to deal with this difficulty and solve (P1) efficiently.

\begin{figure*}[!t]
	\small
	\vspace{-0.25cm}
	\setcounter{equation}{17}
	\begin{subequations}
	\label{eq18}
	\begin{align}
		\frac{\partial f(\mathbf{r}_n)}{\partial x_{r,n}}&=-\frac{2\pi}{\lambda}\sum_{i=1}^{L_r}\sum_{j=1}^{L_r}|b_n^{ij}|\left(\sin\theta_{1,i}\cos\phi_{1,i}-\sin\theta_{1,j}\cos\phi_{1,j}\right)\sin(\Gamma_n^{ij}(\mathbf{r}_n))-\frac{4\pi}{\lambda}\sum_{p=1}^{L_r}|q_n^p|\sin\theta_{1,p}\cos\phi_{1,p}\sin(\kappa_n^p(\mathbf{r}_n)),\\
		\frac{\partial f(\mathbf{r}_n)}{\partial y_{r,n}}&=-\frac{2\pi}{\lambda}\sum_{i=1}^{L_r}\sum_{j=1}^{L_r}|b_n^{ij}|(\cos\theta_{1,i}-\cos\theta_{1,j})\sin(\Gamma_n^{ij}(\mathbf{r}_n))-\frac{4\pi}{\lambda}\sum_{p=1}^{L_r}|q_n^p|\cos\theta_{1,p}\sin(\kappa_n^p(\mathbf{r}_n)).
	\end{align}
	\end{subequations}
	\hrulefill
	\vspace{-0.15cm}
\end{figure*}
\section{Proposed Solution for (P1)}
In this section, we decompose (P1) into several subproblems with respect to $\{\mathbf{W}\}$, $\{\mathbf{r}_n\}_{n=1}^N$ and $\{\mathbf{t}_n\}_{n=1}^N$, respectively, and solve them alternately until convergence.

\subsection{Optimizing $\mathbf{W}$ with Given $\{\mathbf{r}_n\}$ and $\{\mathbf{t}_n\}$}
For any given $\tilde{\mathbf{r}}$ and $\tilde{\mathbf{t}}$, we apply the property of Kronecker product\cite{ref30}, i.e.,
\begin{align}
	\setcounter{equation}{8}
	&\mathrm{vec}(\mathbf{A}_1\mathbf{A}_2\mathbf{A}_3)=\left(\mathbf{A}_3^T\otimes\mathbf{A}_1\right)\cdot\mathrm{vec}(\mathbf{A}_2),\\
	&\left(\mathbf{A}_1\otimes\mathbf{A}_2\right)^T=\mathbf{A}_1^T\otimes\mathbf{A}_2^T,
\end{align}
to re-express (P1) as
\begin{subequations}
	\label{eq11}
	\begin{align}
		\max_{\mathbf{w}}& \quad \frac{P_s\mathbf{w}^H\mathbf{h}\mathbf{h}^H\mathbf{w}}{\sigma_r^2\mathbf{w}^H\mathbf{A}\mathbf{A}^H\mathbf{w}+\sigma_d^2}\\
		\mathrm{s.t.}
		& \quad 
		P_s\mathbf{w}^H\mathbf{B}\mathbf{B}^H\mathbf{w}+\sigma_r^2\mathbf{w}^H\mathbf{w}\leq P_{\mathrm{tot}},
	\end{align}
\end{subequations}
where $\mathbf{w}=\mathrm{vec}(\mathbf{W}),\mathbf{h}=\mathbf{h}_1^*\otimes\mathbf{h}_2,\mathbf{A}=\mathbf{I}\otimes\mathbf{h}_2$ and $\mathbf{B}=\mathbf{h}_1^*\otimes\mathbf{I}$. Although problem \eqref{eq11} is still non-convex, we reveal its hidden convexity by introducing the semidefinite relaxation (SDR). Specifically, let $\mathbf{Q}\triangleq\mathbf{w}\mathbf{w}^H$, problem (11) can be recast as
\begin{subequations}
	\label{eq12}
	\begin{align}
		\max_{\mathbf{Q},\mathrm{rank}(\mathbf{Q})=1}& \quad \frac{\mathrm{Tr}\left(P_s\mathbf{h}\mathbf{h}^H\mathbf{Q}\right)}{\mathrm{Tr}\left(\sigma_r^2\mathbf{A}\mathbf{A}^H\mathbf{Q}\right)+\sigma_d^2}\\
		\mathrm{s.t.}
		& \quad 
		\mathrm{Tr}\left(\left(P_s\mathbf{B}\mathbf{B}^H+\sigma_r^2\mathbf{I}\right)\mathbf{Q}\right)\leq P_{\mathrm{tot}}.
	\end{align}
\end{subequations}
By dropping the rank-one constraint in problem \eqref{eq12}, it becomes a fractional semidefinite programming (SDP) problem, where the objective function is a quasi-affine function of $\mathbf{Q}$. As a result, it can be optimally solved by combining the interior-point algorithm and bisection search. However, we show that it can be further recast into a linear programming problem and solved in a more efficient manner. To this end, we introduce an auxiliary variable $\tau \ge 0$ and let $\mathbf{Q}=\tilde{\mathbf{Q}}/\tau$ with $\tilde{\mathbf{Q}}\succeq\mathbf{0}$. By invoking the Charnes-Cooper transformation\cite{ref31}, problem \eqref{eq12} can be equivalently recast as
\begin{subequations}
	\label{eq13}
	\begin{align}
		\min_{\tilde{\mathbf{Q}},\tau}& \quad \mathrm{Tr}\left(\sigma_r^2\mathbf{A}\mathbf{A}^H\tilde{\mathbf{Q}}\right)+\tau\sigma_d^2\\
		\mathrm{s.t.}
		& \quad 
		\mathrm{Tr}\left(\left(P_s\mathbf{B}\mathbf{B}^H+\sigma_r^2\mathbf{I}\right)\tilde{\mathbf{Q}}\right)\leq \tau P_{\mathrm{tot}},\\
		& \quad \mathrm{Tr}\left(P_s\mathbf{h}\mathbf{h}^H\tilde{\mathbf{Q}}\right)=1,~\tilde{\mathbf{Q}}\succeq\mathbf{0},
	\end{align}
\end{subequations}
which is a semidefinite programming (SDP) problem and thus can be optimally solved via the interior-point algorithm. Let $\{\tilde{\mathbf{Q}}^{\star},\tau^{\star}\}$ denote the optimal solution to problem  $\eqref{eq13}$. Then, the optimal solution to problem \eqref{eq12} can be retrieved as $\mathbf{Q}^{\star}=\tilde{\mathbf{Q}}^{\star}/\tau^{\star}$. Next, we present the following proposition to show that $\mathrm{rank}(\mathbf{Q}^{\star})=1$, such that we can always reconstruct a rank-one optimal solution to problem (11), i.e., the SDR is always tight.

\textit{Proposition 1}: If problem \eqref{eq12} is feasible, then its optimal solution $\mathbf{Q}^{\star}$ must satisfies $\mathrm{rank}(\mathbf{Q}^{\star})=1$.

\textit{Proof}: See Appendix A.$\hfill\blacksquare$

Based on Proposition 1, we can obtain an optimal solution $\mathbf{w}^{\star}$ from $\mathbf{Q}^{\star}$ through eigenvalue decomposition. Specifically, denote by $\lambda_{\max}$ and $\mathbf{x}$ the maximum eigenvalue of $\mathbf{Q}^{\star}$ and the corresponding eigenvector, respectively. Then, the optimal solution to problem \eqref{eq11} can be constructed as $\mathbf{w}^{\star}=\sqrt{\lambda_{\max}}\mathbf{x}$. 

\subsection{Optimizing $\{\mathbf{r}_n\}_{n=1}^N$ with Given $\mathbf{W}$ and $\{\mathbf{t}_n\}$}
First, we define $\mathbf{h}_1(\tilde{\mathbf{r}})=[h_{1,1}(\mathbf{r}_1),\cdots,h_{1,N}(\mathbf{r}_N)]^T, \mathbf{W}=[\mathbf{w}_1,\cdots,\mathbf{w}_N]$, and $\mathbf{a}=\mathbf{W}^H\mathbf{h}_2=[a_1,\cdots,a_N]^T$. Then, for any given $\mathbf{W},\tilde{\mathbf{t}}$ and $\{\mathbf{r}_m\}_{m\neq n}$, (P1) can be simplified as
\begin{subequations}
	\label{eq14}
	\begin{align}
		\max_{\mathbf{r}_n}& \quad \left|a_n^*h_{1,n}(\mathbf{r}_n)+\alpha_n\right|^2\label{eq14a}\\
		\mathrm{s.t.}
		& \quad 
		P_s\|h_{1,n}(\mathbf{r}_n)\mathbf{w}_n+\mathbf{b}_n\|^2\leq \tilde{P}_{\mathrm{tot}},\\
		& \quad \eqref{eq8b},~\eqref{eq8c},\notag
	\end{align}
\end{subequations}
where $\alpha_n=\sum_{k\neq n}a_k^*h_{1,k}(\mathbf{r}_k),\mathbf{b}_n=\sum_{k\neq n}h_{1,k}(\mathbf{r}_k)\mathbf{w}_k$, and $\tilde{P}_{\mathrm{tot}}=P_{\mathrm{tot}}-\sigma_r^2\|\mathbf{W}\|^2$. Note that $\alpha_n$ is independent of $\mathbf{r}_n$. Thus, maximizing \eqref{eq14a} is equivalent to maximizing
\begin{align}
	\label{eq15}
	f(\mathbf{r}_n)&=|a_n|^2|h_{1,n}(\mathbf{r_n})|^2+2\mathrm{Re} \left\{a_n\alpha_nh_{1,n}^*(\mathbf{r}_n)\right\} \notag\\
	&=\underbrace{\mathbf{f}_1^H(\mathbf{r}_n)\mathbf{B}_n\mathbf{f}_1(\mathbf{r}_n)}_{f_1(\mathbf{r}_n)}+2\underbrace{\mathrm{Re}\left\{\mathbf{q}_n^H\mathbf{f}_1(\mathbf{r}_n)\right\}}_{f_2(\mathbf{r}_n)},
\end{align}
where $\mathbf{B}_n=|a_n|^2\mathbf{g}_1\mathbf{g}_1^H$ and $\mathbf{q}_n=a_n^*\alpha_n^*\mathbf{g}_1$. However, \eqref{eq15} is still a non-concave function with respect to (w.r.t.) $\mathbf{r}_n$. To tackle this problem, we use the GA method to obtain a locally optimal solution. To this end, we first derive the gradient vector of \eqref{eq15}, i.e., $\nabla_{\mathbf{r}_n} f(\mathbf{r}_n)$. By denoting the $(i,j)$-th entry of $\mathbf{B}_n$ as $b_n^{ij}=|b_n^{ij}|e^{\mathrm{j}\angle b_n^{ij}}$, with $|b_n^{ij}|$ and $\angle b_n^{ij}$ representing its amplitude and phase, respectively, we can expand $f_1(\mathbf{r}_n)$ as
\begin{align}
	\label{eq16}
	f_1(\mathbf{r}_n)&=\mathrm{Re}\left\{\sum_{i=1}^{L_r}\sum_{j=1}^{L_r}|b_n^{ij}|e^{\mathrm{j}\left(\frac{2\pi}{\lambda}(-\rho_{1,i}(\mathbf{r}_n)+\rho_{1,j}(\mathbf{r}_n))+\angle b_n^{ij}\right)}\right\}\notag\\
	&=\sum_{i=1}^{L_r}\sum_{j=1}^{L_r}|b_n^{ij}|\cos (\Gamma_n^{ij}(\mathbf{r}_n)),
\end{align}
with $\Gamma_n^{ij}(\mathbf{r}_n)=\frac{2\pi}{\lambda}(\rho_{1,i}(\mathbf{r}_n)-\rho_{1,j}(\mathbf{r}_n))-\angle b_n^{ij}$. Let $q_n^p=|q_n^p|e^{\mathrm{j}\angle q_n^p}$ denote the $p$-th entry of $\mathbf{q}_n$ with $|q_n^p|$ and $\angle q_n^p$ being its amplitude and phase, respectively. Similarly to \eqref{eq16}, we can expand $f_2(\mathbf{r}_n)$ as
\begin{align}
	f_2(\mathbf{r}_n)&=\mathrm{Re}\left\{\sum_{p=1}^{L_r}|q_n^{p}|e^{\mathrm{j}\left(\frac{2\pi}{\lambda}\rho_{1,p}(\mathbf{r}_n)-\angle q_n^p\right)}\right\}\notag\\
	&=\sum_{p=1}^{L_r}|q_n^p|\cos (\kappa_n^p(\mathbf{r}_n))
\end{align}
with $\kappa_n^p(\mathbf{r}_n)=\frac{2\pi}{\lambda}\rho_{1,p}(\mathbf{r}_n)-\angle q_n^p$. It follows from the above that the gradient of \eqref{eq15} can be derived as $\nabla_{\mathbf{r}_n} f(\mathbf{r}_n)=\left[\frac{\partial f(\mathbf{r}_n)}{\partial x_{r,n}},\frac{\partial f(\mathbf{r}_n)}{\partial y_{r,n}}\right]^T$, as shown in \eqref{eq18}. Then, based on the principle of the GA, we can update $\mathbf{r}_n$ in the GA iterations as
\begin{align}
	\setcounter{equation}{18}
	\mathbf{r}_n^{s+1}=\mathbf{r}_n^{s}+\mu^s \nabla_{\mathbf{r}_n^s} f(\mathbf{r}_n^s),
\end{align}
where $s$ denotes the iteration number and $\mu^s$ denotes the step size of the GA in the $s$-th iteration. To ensure that each $\mathbf{r}_n^s$ satisfies the constraints in problem \eqref{eq14}, we need to adjust the step size in each iteration. Define $\mathcal{R}_n=\{\mathbf{r}_n|\mathbf{r}_n\in\mathcal{C}_r;\| \mathbf{r}_m-\mathbf{r}_n\|_2\geq D,\forall m\neq n;P_s\|h_{1,n}(\mathbf{r}_n)\mathbf{w}_n+\mathbf{b}_n\|^2\leq \tilde{P}_{\mathrm{tot}}\}$ as the feasible set of $\mathbf{r}_n$. Then, in the $(s+1)$-th iteration, if $\mathbf{r}_n^{s+1}$ is not within the fesible set, i.e., $\mathbf{r}_n^{s+1}\notin\mathcal{R}_n$, the step size is reset as $\mu^s=\mu^s/2$. This process is repeated until $\mathbf{r}_n^{s+1}\in\mathcal{R}_n$. The procedures of our proposed GA-based algorithm for optimizing $\{\mathbf{r}_n\}_{n=1}^N$ are summarized in \textbf{Algorithm \ref{alg1}}.
\begin{figure}[t]
\vspace{-0.25cm}
\begin{algorithm}[H]
	\footnotesize
	\caption{GA-based algorithm for optimizing $\tilde{\mathbf{r}}$}
	\begin{algorithmic}[1]
		\STATE \textbf{Input} $\tilde{\mathbf{r}}^0=[\mathbf{r}_1^0,\cdots,\mathbf{r}_N^0]$, $s=0$, $\mu_{\mathrm{ini}}$
		\STATE \textbf{repeat}
		
		\STATE \hspace{0.25cm} \textbf{for} $n=1\rightarrow N$ \textbf{do}
		
		\STATE \hspace{0.5cm} Calculate $\nabla_{\mathbf{r}_n^s} f(\mathbf{r}_n)$ via \eqref{eq18} and set $\mu^s=\mu_{\mathrm{ini}}$.
		
		\STATE \hspace{0.5cm} Update $\hat{\mathbf{r}}_n=\mathbf{r}_n^{s}+\mu^s \nabla_{\mathbf{r}_n^s} f(\mathbf{r}_n^s)$.

		\STATE \hspace{0.5cm} \textbf{while} $f(\hat{\mathbf{r}}_n)\notin\mathcal{R}_n$ or $f(\hat{\mathbf{r}}_n)<f(\mathbf{r}_n^s)$
		
		\STATE \hspace{0.75cm} Set $\mu^s=\mu^s/2$ and update $\hat{\mathbf{r}}_n=\mathbf{r}_n^{s}+\mu^s \nabla_{\mathbf{r}_n^s} f(\mathbf{r}_n^s)$.
		
		\STATE \hspace{0.5cm} \textbf{end while}
		
		\STATE \hspace{0.5cm}  Update $\mathbf{r}_n^{s+1}=\hat{\mathbf{r}}_n$.
		
		\STATE \hspace{0.25cm} \textbf{end for}
		
		\STATE \hspace{0.25cm} $s=s+1$.
		
		\STATE \textbf{until} the objective of \eqref{eq14}  converges to a prescribed
		accuracy.
		
	\end{algorithmic}
	\label{alg1} 
\end{algorithm}
\vspace{-0.5cm}
\end{figure}

\subsection{Optimizing $\{\mathbf{t}_n\}_{n=1}^N$ with Given $\mathbf{W}$ and $\{\mathbf{r}_n\}$}
Due to the monotonically increasing property of the logarithmic function, for any given $\mathbf{W},\tilde{\mathbf{t}}$ and $\{\mathbf{t}_m\}_{m\neq n}$, (P1) can be simplified as
\begin{subequations}
	\label{eq20}
	\begin{align}
		\max_{\mathbf{t}_n}& \quad g(\mathbf{t}_n)=\log_2\left(P_s\left|\mathbf{h}_2^H(\tilde{\mathbf{t}})\mathbf{W}\mathbf{h}_1\right|^2\right)\notag\\
		& \qquad \qquad~ -\log_2\left(\sigma_r^2\left\|\mathbf{h}_2^H(\tilde{\mathbf{t}})\mathbf{W}\right\|^2+\sigma_d^2\right)\\
		\mathrm{s.t.}
		& \quad \eqref{eq8b},~\eqref{eq8d}.\notag
	\end{align}
\end{subequations}
Similarly to Section III-B, we ultilize the GA method to solve problem $\eqref{eq20}$. Define $\mathbf{h}_2(\tilde{\mathbf{t}})=[h_{2,1}(\mathbf{t}_1),\cdots,h_{2,N}(\mathbf{t}_N)]^T$, $\mathbf{W}^H=[\tilde{\mathbf{w}}_1,\cdots,\tilde{\mathbf{w}}_N]$, and $\mathbf{c}=\mathbf{W}\mathbf{h}_1=[c_1,\cdots,c_N]^T$. Then, we can rewrite the objective of problem \eqref{eq20} as
\begin{align}
	g(\mathbf{t}_n)=&\underbrace{\log_2\left(P_s\left|c_n^*h_{2,n}(\mathbf{t}_n)+\beta_n\right|^2\right)}_{g_1(\mathbf{t}_n)}\notag\\
	&-\underbrace{\log_2\left(\sigma_r^2\left\|h_{2,n}(\mathbf{t}_n)\tilde{\mathbf{w}}_n+\mathbf{d}_n\right\|^2+\sigma_d^2\right)}_{g_2(\mathbf{t}_n)},
\end{align}
where $\beta_n=\sum_{k\neq n}c_k^*h_{2,k}(\mathbf{t}_k)$ and $\mathbf{d}_n=\sum_{k\neq n}h_{2,k}(\mathbf{t}_k)\tilde{\mathbf{w}}_k$. Note that $\beta_n$ and $\mathbf{d}_n$ are independent of $\mathbf{t}_n$, based on which we can expand $g_1(\mathbf{t}_n)$ as
\begin{equation}
	g_1(\mathbf{t}_n)=\log_2\left(g_{1,1}(\mathbf{t}_n)+2g_{1,2}(\mathbf{t}_n)+\mathrm{constant}\right)
\end{equation}
with
\begin{align}
	g_{1,1}(\mathbf{t}_n)&=P_s|c_n|^2|h_{2,n}(\mathbf{t}_n)|^2=\mathbf{g}_2^H(\mathbf{t}_n)\mathbf{E}_n\mathbf{g}_2(\mathbf{t}_n)\\
	g_{1,2}(\mathbf{t}_n)&=\mathrm{Re}\left\{P_sc_n\beta_nh_{2,n}^*(\mathbf{t}_n)\right\}=\mathrm{Re}\left\{\mathbf{m}_n^H\mathbf{g}_2(\mathbf{t}_n)\right\}
\end{align}
where $\mathbf{E}_n=P_s|c_n|^2\mathbf{f}_2\mathbf{f}_2^H$ and $\mathbf{m}_n=P_sc_n^*\beta_n^*\mathbf{f}_2$. Following the derivation of \eqref{eq18}, we can also derive the gradient of $g_{1,1}(\mathbf{t}_n)$ and $g_{1,2}(\mathbf{t}_n)$ w.r.t. $\mathbf{t}_n$, i.e., $\nabla_{\mathbf{t}_n}g_{1,1}(\mathbf{t}_n)$ and $\nabla_{\mathbf{t}_n}g_{1,2}(\mathbf{t}_n)$, which result in
\begin{equation}
	\label{eq26}
	\nabla_{\mathbf{t}_n}g_1(\mathbf{t}_n)=\frac{\nabla_{\mathbf{t}_n}g_{1,1}(\mathbf{t}_n)+2\nabla_{\mathbf{t}_n}g_{1,2}(\mathbf{t}_n)}{P_s\left|c_n^*h_{2,n}(\mathbf{t}_n)+\beta_n\right|^2}.
\end{equation}
Similarly, we can expand $g_2(\mathbf{t}_n)$ as
\begin{equation}
	g_2(\mathbf{t}_n)=\log_2\left(g_{2,1}(\mathbf{t}_n)+2g_{2,2}(\mathbf{t}_n)+\mathrm{constant}\right)
\end{equation}
with
\begin{align}
	g_{2,1}(\mathbf{t}_n)&=\sigma_r^2\|\tilde{\mathbf{w}}_nh_{2,n}(\mathbf{t}_n)\|^2=\mathbf{g}_2^H(\mathbf{t}_n)\mathbf{F}_n\mathbf{g}_2(\mathbf{t}_n)\\
	g_{2,2}(\mathbf{t}_n)&=\mathrm{Re}\left\{\tilde{\mathbf{w}}_n^H\mathbf{d}_nh_{2,n}^*(\mathbf{t}_n)\right\}=\mathrm{Re}\left\{\mathbf{s}_n^H\mathbf{g}_2(\mathbf{t}_n)\right\}
\end{align}
where $\mathbf{F}_n=\sigma_r^2\|\tilde{\mathbf{w}}_n\|^2\mathbf{f}_2\mathbf{f}_2^H$ and $\mathbf{s}_n=\mathbf{f}_2\mathbf{d}_n^H\tilde{\mathbf{w}}_n$. Then, the gradient of $g_{2,1}(\mathbf{t}_n)$ and $g_{2,2}(\mathbf{t}_n)$ w.r.t. $\mathbf{t}_n$, i.e., $\nabla_{\mathbf{t}_n}g_{2,1}(\mathbf{t}_n)$ and $\nabla_{\mathbf{t}_n}g_{2,2}(\mathbf{t}_n)$, can be derived similarly, which result in
\begin{equation}
	\label{eq30}
	\nabla_{\mathbf{t}_n}g_2(\mathbf{t}_n)=\frac{\nabla_{\mathbf{t}_n}g_{2,1}(\mathbf{t}_n)+2\nabla_{\mathbf{t}_n}g_{2,2}(\mathbf{t}_n)}{\sigma_r^2\left\|\mathbf{h}_2^H(\tilde{\mathbf{t}})\mathbf{W}\right\|^2+\sigma_d^2}.
\end{equation}
Finally, by combining \eqref{eq26} and \eqref{eq30}, the gradient of the obejctive function of problem \eqref{eq20} can be calculated as
\begin{equation}
	\nabla_{\mathbf{t}_n}g(\mathbf{t}_n)=\nabla_{\mathbf{t}_n}g_1(\mathbf{t}_n)-\nabla_{\mathbf{t}_n}g_2(\mathbf{t}_n).
\end{equation}
Therefore, the update rule for $\mathbf{t}_n$ is given by
\begin{equation}
	\label{eq32}
	\mathbf{t}_n^{s+1}=\mathbf{t}_n^s+\mu^s\nabla_{\mathbf{t}_n^s}g(\mathbf{t}_n^s).
\end{equation}
Define $\mathcal{T}_n=\{\mathbf{t}_n|\mathbf{t}_n\in\mathcal{C}_r;\| \mathbf{t}_m-\mathbf{t}_n\|_2\geq D,\forall m\neq n\}$ as the feasible set of $\mathbf{t}_n$. To ensure that each $\mathbf{t}_n^s$ satisfies the constraints in problem \eqref{eq20}, we dynamically adjust the step size in each iteration, similarly to the optimization of $\mathbf{r}_n$. The procedures of our proposed GA-based algorithm for optimizing $\{\mathbf{t}_n\}_{n=1}^N$ resemble \textbf{Algorithm \ref{alg1}}, by simply replacing $\left\{\tilde{\mathbf{r}}^0=[\mathbf{r}_1^0,\cdots,\mathbf{r}_N^0],~f(\mathbf{r}_n),~\nabla_{\mathbf{r}_n}f(\mathbf{r}_n),~\mathcal{R}_n\right\}$ therein with $\left\{\tilde{\mathbf{t}}^0=[\mathbf{t}_1^0,\cdots,\mathbf{t}_N^0],~g(\mathbf{t}_n),~\nabla_{\mathbf{t}_n}g(\mathbf{t}_n),~\mathcal{T}_n\right\}$.

\subsection{Overall Algorithm}
The overall AO algorithm for solving (P1) is summarized in \textbf{Algorithm \ref{alg2}}. Notably, the convergence of this algorithm is always guaranteed since the objective is non-decreasing over iterations and has an upper bound. 
\begin{algorithm}[!h]
	\footnotesize
	\caption{AO algorithm for solving problem \eqref{eq8}}
	\begin{algorithmic}[1]
		\STATE \textbf{Initialize} $\mathbf{W}$, $\tilde{\mathbf{r}}$, $\tilde{\mathbf{t}}$
		\STATE \textbf{repeat}
		
		\STATE \hspace{0.25cm} Set $\tilde{\mathbf{r}}^0=\tilde{\mathbf{r}}$ and update $\tilde{\mathbf{r}}$ via Algorithm 1.
		
		\STATE \hspace{0.25cm} Set $\tilde{\mathbf{t}}^0=\tilde{\mathbf{t}}$ and update $\tilde{\mathbf{t}}$ similarly to Algorithm 1.				
		
		\STATE \hspace{0.25cm} Obtain $\{\tilde{\mathbf{Q}},\tau\}$ by solving problem \eqref{eq13} and set $\mathbf{Q}=\tilde{\mathbf{Q}}/\tau$.
		
		\STATE \hspace{0.25cm} Obtain the maximum eigenvalue $\lambda_{\max}$ of $\mathbf{Q}$ and its corresp-\\
		\hspace{0.25cm}  onding eigenvector $\mathbf{x}$ via eigenvalue decomposition.
		
		\STATE \hspace{0.25cm} Set $\mathbf{w}=\sqrt{\lambda_{\max}}\mathbf{x}$ and update $\mathbf{W}=\mathrm{vec}^{-1}(\mathbf{w})$.
		
		\STATE \textbf{until} convergence
		
	\end{algorithmic}
	\label{alg2} 
\end{algorithm}

\textit{Complexity Analysis}: The complexity for optimizing $\mathbf{W}$ is mainly from solving the SDP problem, i.e., problem \eqref{eq13}, whose complexity is $\mathcal{O}(N^7)$\cite{ref33}. The complexity for optimizing $\tilde{\mathbf{r}}$ and $\tilde{\mathbf{t}}$ is given by $\mathcal{O}(I_1NL_r^2)$ and $\mathcal{O}(I_2NL_t^2)$, respectively, where $I_1$ and $I_2$ denote the maximum GA iteration numbers. Hence, the overall complexity of our proposed algorithm is $\mathcal{O}\left(T\left(N^7+I_1NL_r^2+I_2NL_t^2\right)\right)$, where $T$ denotes the maximum AO iteration number.

\section{Simulation Results}

\begin{figure}[!t]
	\vspace{-0.25cm}
	\centering
	\includegraphics[width=0.39\textwidth]{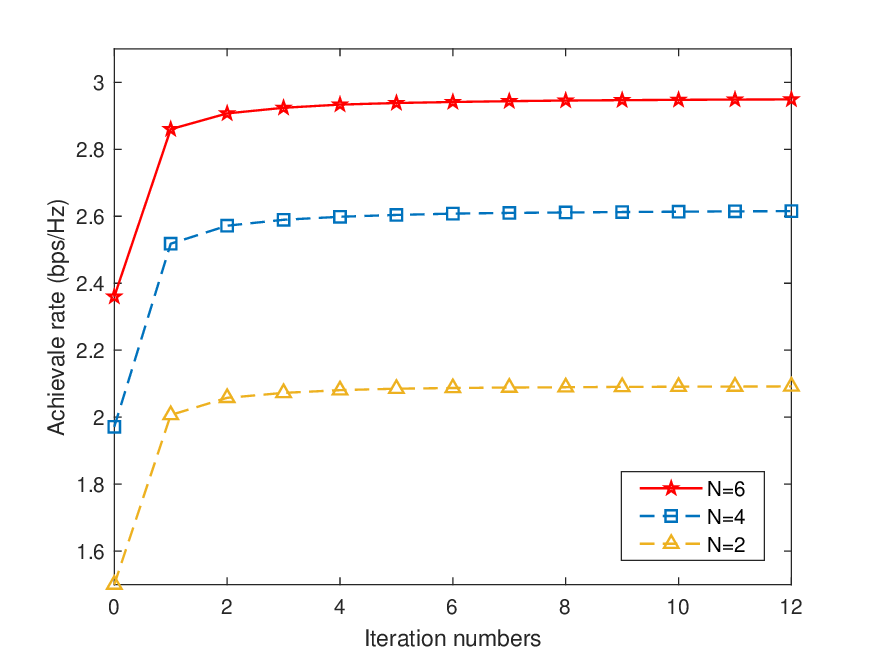}
	\caption{Convergence plot of our proposed AO algorithm.}
	\label{convergence_performance}
	\vspace{-0.25cm}
\end{figure}
\begin{figure}[t]
	\centering
	\includegraphics[width=0.39\textwidth]{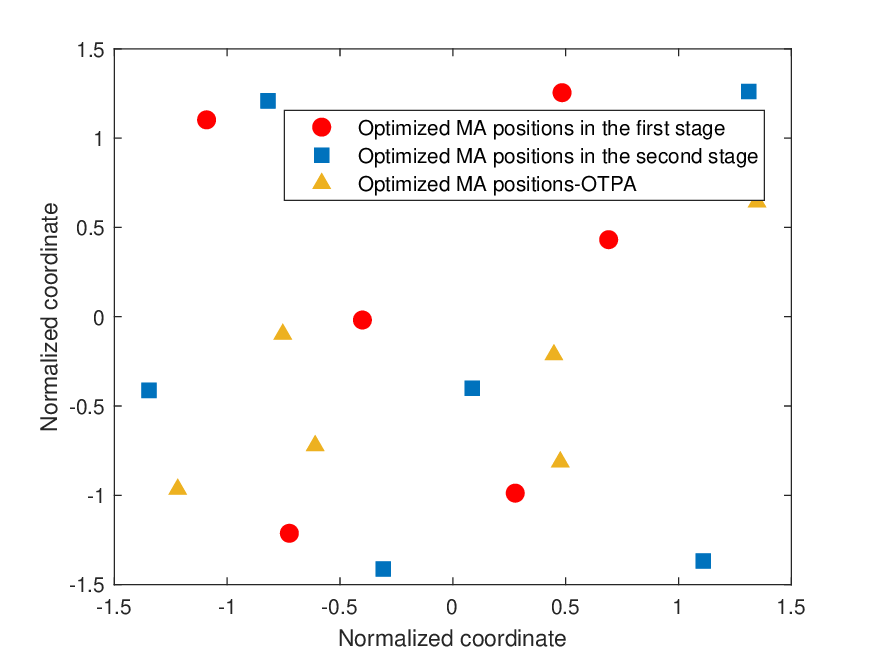}
	\caption{Optimized positions of the MAs at the relay.}
	\label{MA_positions}
	\vspace{-0.35cm}
\end{figure}
In this section, we present simulation results to show the effectiveness of our proposed MA-assisted relay system. In the simulation, the PRV from the source to the relay and that from the relay to the detination are both modeled as circularly symmetric complex Gaussian random vectors with independent and identically distributed (i.i.d.) elements, i.e., $g_{1,i}\sim\mathcal{CN}(0,1/L_r)$,$f_{2,i}\sim\mathcal{CN}(0,1/L_t)$. The elevation
and azimuth AoDs/AoAs are modeled as i.i.d. uniformly distributed variables over $[0,2\pi]$. Besides, the number of the receive paths from the source to the relay is set the same as that from the relay to the destination, i.e., $L_r=L_t=5$, the noise power at the relay/destination is set to $\sigma_r^2=\sigma_d^2=1$, and the minimum inter-MA distance is set to $D=\lambda/2$. All the results are averaged over 1000 independent channel realizations. 

\begin{figure}[t]
	\vspace{-0.25cm}
	\centering
	\includegraphics[width=0.39\textwidth]{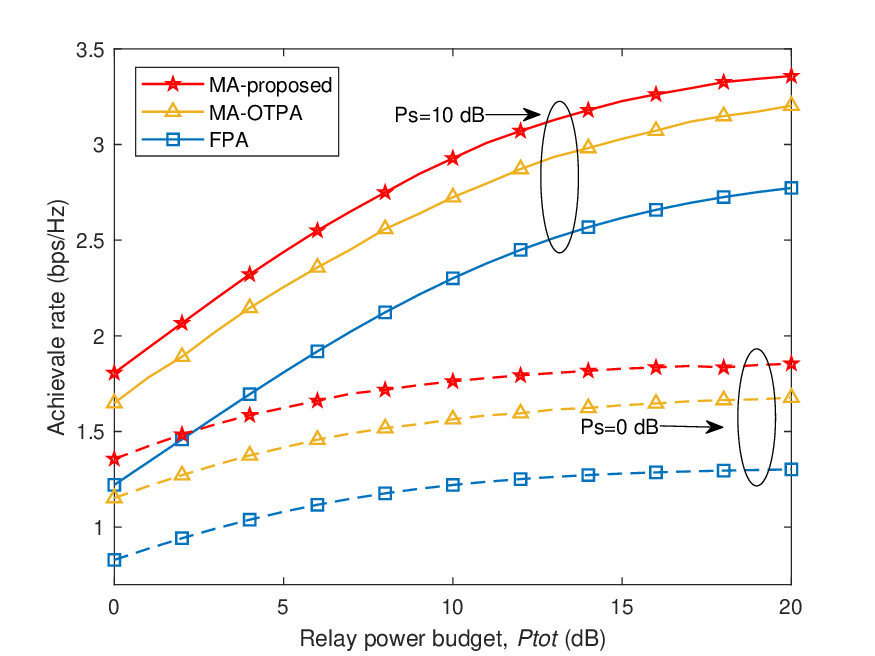}
	\caption{Achievable rates versus the relay power budget.}
	\label{rate_vs_P_tot}
	\vspace{-0.25cm}
\end{figure}

\begin{figure}[t]
	\centering
	\includegraphics[width=0.39\textwidth]{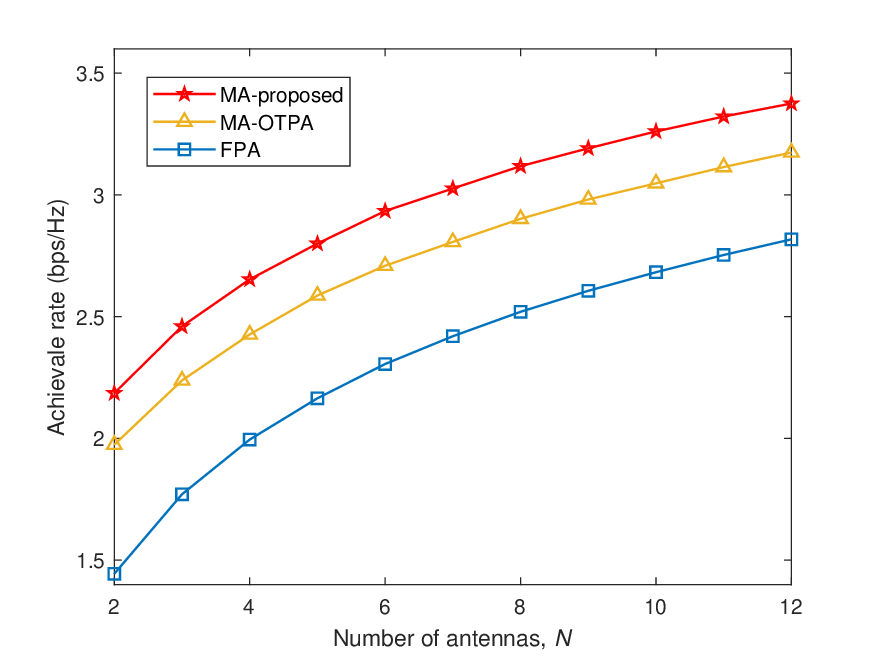}
	\caption{Achievable rates versus the number of antennas.}
	\label{rate_vs_N}
	\vspace{-0.35cm}
\end{figure}

For performance comparison, we consider the following benchmark schemes: \textbf{1) FPA}: The relay is equipped with an FPA-based uniform planar array with $N$ antennas and the spacing between any two adjacent antennas is set to $\lambda/2$; \textbf{2) One-time position adjustment (OTPA)}: The positions of the MAs at the relay are adjusted only once to cater to both the reception/transmission from/to the source/destination. The associated antenna position optimization problem can also be solved by applying the GA algorithm, for which the details are omitted for brevity.

In Fig. \ref{convergence_performance}, we show the convergence behavior of the achievable rate at the destination (i.e., $\frac{1}{2}\log_2(1+\gamma)$ in bps/Hz, where  ``1/2'' is due to the half-duplex processing of the relay) by our proposed AO algorithm, with $A=3\lambda$ and $P_s=P_{\mathrm{tot}}=10~\mathrm{dB}$. It is observed that our proposed AO algorithm converges  after 10 iterations for all values of $N$ considered, which is consistent with our theoretical analysis in Section III-D.

Fig. \ref{MA_positions} shows the optimized MA positions by our proposed AO algorithm and the OTPA scheme with $N=6$. First, it is observed that the positions of the MAs in both schemes are not arranged in a regular manner as the conventional FPAs, in order to achieve a better channel condition. Besides, there exists significant differences between the MA positions by these two schemes, as the OTPA scheme needs to accommodate both the signal reception and transmission of the relay.

Fig. \ref{rate_vs_P_tot} shows the achievable rate versus the total power budget of the relay, with $N=6$ and $A=3\lambda$. It is observed that our proposed scheme achieves the highest rate among all considered schemes for both $P_s=0~\mathrm{dB}$ and $P_s=10~\mathrm{dB}$. Nonetheless, the OTPA scheme can achieve a small performance gap with our proposed scheme, implying that the performance loss by only adjusting the MA positions once may not be significant under certain conditions. Moreover, it can be found that the achievable rate increases faster with the relay power budget when the transmit power at the source is large. This is expected, as a low transmit power budget limits the amplification gain by the relay and thus the achievable rate performance.

\begin{figure}[t]
	\vspace{-0.25cm}
	\centering
	\includegraphics[width=0.39\textwidth]{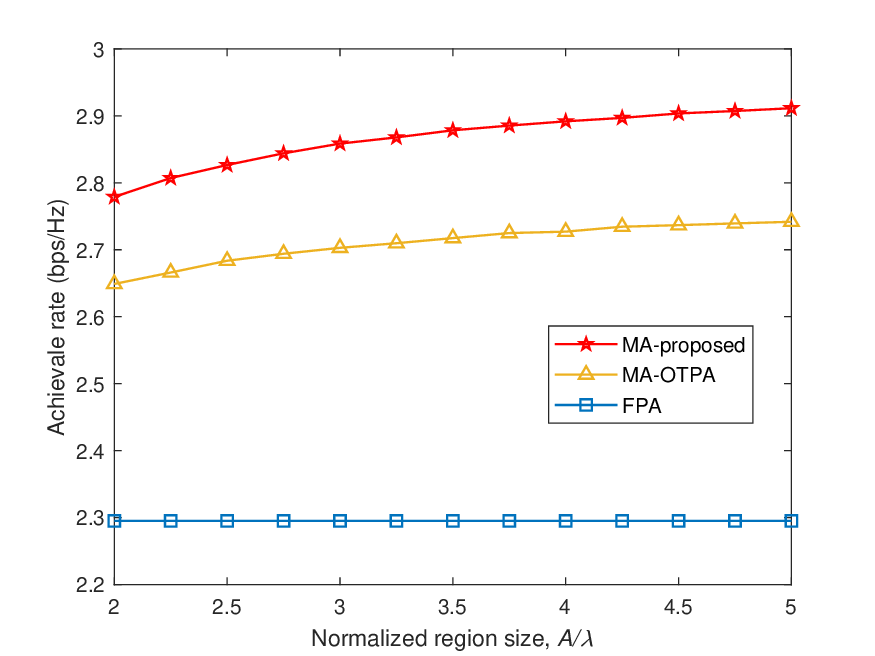}
	\caption{Achievable rates versus the normalized region size.}
	\label{rate_vs_A}
	\vspace{-0.4cm}
\end{figure}
Fig. \ref{rate_vs_N} shows the achievable rate versus the number of antennas with $A=3\lambda$ and $P_s=P_{\mathrm{tot}}=10~\mathrm{dB}$. It is observed that our proposed scheme outperforms other benchmark schemes under different value of $N$. Besides, with increasing the antenna numbers, all considered schemes can achieve a higher date rate thanks to the enhanced spatial diversity gain and beamforming gain. The OTPA scheme is observed to achieve a close performance to the proposed scheme (less than 0.2 bps/Hz). Finally, in Fig. \ref{rate_vs_A}, we plot the achievable rate versus the normalized region size, i.e., $A/\lambda$, with $N=6$ and $P_s=P_{\mathrm{tot}}=10~\mathrm{dB}$. It is observed that the achievable rate increases with the region size in the proposed scheme and the OTPA scheme, as it enables MAs to enjoy more available spatial diversity gain. However, when the region is sufficiently large, the spatial diversity gain will saturate. Accordingly, the achievable rates by these two schemes finally converge. It is also observed that the achievable rate by the proposed scheme increases faster than that by the OTPA scheme, since more spatial diversity gain is exploited in the former scheme with two-stage (versus one-stage) antenna position optimization.

\section{Conclusion}
This paper studied a joint beamforming and two-stage antenna position optimization problem for an MA-enhanced AF relaying system, aiming to maximize the achievable rate at the destination.  To deal with the non-convexity due to the two-stage antenna position optimization, we proposed an AO algorithm to decompose the primal problem into several subproblems and solve them separately by combining the SDP and the GA algorithms. Simulation results showed that our proposed system can significantly enhance the AF relaying performance compared with the conventional FPA system. It was also shown that the OTPA scheme may achieve a close performance to the two-stage antenna position optimization.

{\appendices
	\section{Proof of Proposition 1}
	Since problem \eqref{eq13} is convex and satisfies Slater’s condition, the duality gap is zero, and the optimal primal/dual solutions must satisfy the Karush-Kuhn-Tucker (KKT) conditions\cite{ref32}. Let $\lambda^{\star}\geq 0,\nu^{\star}$ and $\mathbf{Y}^{\star}\succeq\mathbf{0}$ denote the optimal dual variables associated with the constraints in problem \eqref{eq13}. According to the KKT conditions, we have:
	\begin{subequations}
		\begin{align}
			&\tilde{\mathbf{Q}}^{\star}\succeq\mathbf{0},\mathbf{Y}^{\star}\succeq\mathbf{0},\lambda^{\star}\geq 0\label{eq33a}\\
			&\mathbf{Y}^{\star}\tilde{\mathbf{Q}}^{\star}=\mathbf{0}\label{eq33b}\\
			&\mathbf{Y}^{\star}=\sigma_r^2\mathbf{A}\mathbf{A}^H+\lambda^{\star}\left(P_s\mathbf{B}\mathbf{B}^H+\sigma_r^2\mathbf{I}\right)-\nu^{\star}P_s\mathbf{h}\mathbf{h}^H.\label{eq33c}
		\end{align}
	\end{subequations}
	Define $\mathbf{R}=\sigma_r^2\mathbf{A}\mathbf{A}^H+P_s\mathbf{B}\mathbf{B}^H+\sigma_r^2\mathbf{I}$. It is easy to verify that $\mathbf{R}$ is a positive definite matrix, which implies that it must be of full rank, i.e., $\mathrm{rank}(\mathbf{R})=N^2$. Substituting \eqref{eq33c} into \eqref{eq33b}, we can obtain the following equality:
	\begin{align}
		\mathbf{R}\tilde{\mathbf{Q}}^{\star}=\nu^{\star}P_s\mathbf{h}\mathbf{h}^H\tilde{\mathbf{Q}}^{\star}.
	\end{align}
	Moreover, since $\mathbf{R}$ is of full rank, we have
	\begin{align}
		\mathrm{rank}(\mathbf{R}\tilde{\mathbf{Q}}^{\star})=\mathrm{rank}(\tilde{\mathbf{Q}}^{\star})&=\mathrm{rank}(\mathbf{h}\mathbf{h}^H\tilde{\mathbf{Q}}^{\star})\notag\\
		&\leq \mathrm{rank}(\mathbf{h}\mathbf{h}^H)=1.
	\end{align}
	Note that $\mathrm{rank}(\tilde{\mathbf{Q}}^{\star})=0$ results in $\tilde{\mathbf{Q}}^{\star}=\mathbf{0}$, which is infeasible to problem \eqref{eq13}. As a result, we must have $\mathrm{rank}(\tilde{\mathbf{Q}}^{\star})=1$, and the optimal solution to problem \eqref{eq12}, i.e., $\mathbf{Q}^{\star}=\tilde{\mathbf{Q}}^{\star}/\tau^{\star}$, must satisfy $\mathrm{rank}(\mathbf{Q}^{\star})=1$. The proof is thus complete.

}

\bibliography{reference}
\bibliographystyle{IEEEtran}

\end{document}